**Eliminating the Perovskite Solar Cell Manufacturing Bottleneck via High-Speed Flexography**

*Julia E. Huddy, Youxiong Ye, and William J. Scheideler\**

Thayer School of Engineering
Dartmouth College, Hanover, NH 03755, United States
E-mail: william.j.scheideler@dartmouth.edu

**Abstract**

Perovskite solar cells have potential to deliver terawatt-scale power via low-cost manufacturing. However, scaling is limited by slow, high-temperature annealing of the inorganic transport layers and the lack of reliable, large-area methods for depositing thin (< 30 nm) charge transport layers (CTLs). We present a method for scaling ultrathin $NiO_x$ hole transport layers (HTLs) by pairing high-speed (60 m/min) flexographic printing with rapidly annealed sol-gel inks to achieve the fastest reported process for fabrication of inorganic CTLs for perovskites. By engineering precursor rheology for rapid film-leveling, $NiO_x$ HTLs were printed with high uniformity and ultralow pinhole densities resulting in photovoltaic performance exceeding that of spin-coated devices. Integrating these printed transport layers in planar inverted PSCs allows rapid fabrication of high-efficiency (PCE > 15%) $Cs_xFA_{1-x}PbI$ solar cells with improved short circuit currents ($J_{sc}$) of 22.4 mA/cm$^2$. Rapid annealing of the HTL accelerates total processing time by 60X, while maintaining the required balance of optoelectronic properties and the chemical composition for effective hole collection. These results build an improved understanding of ultrathin $NiO_x$ and reveal opportunities to enhance device performance via scalable manufacturing of inorganic CTLs.

# 1. Introduction

With their fundamental material advantages including slow radiative recombination,[1] high absorption coefficients,[2] and amenability to low-cost solution-deposition,[3] metal halide perovskites (PVSKs) could become the first thin film solar technology to deliver on the promise of terawatt-scale photovoltaic (PV) capacity. Deeper global PV integration demands further cost reductions and new functionality beyond capabilities of rigid Si panels.[4] For example, opportunities for solar integration with electric vehicles[5] demand non-standard module geometries for which c-silicon is unfit and current generations of flexible solar cells (a-Si) are too low in efficiency. However, before lightweight, flexible perovskites can enable deeper renewable integration, fundamental scientific and technological advances in scalable manufacturing must overcome the photovoltaic performance and reliability drop off exhibited by large area devices (10 – 100 cm$^2$).[6]

Recent advances in rapid manufacturing of the PVSK absorber by blade coating,[7] spray coating,[8] inkjet printing,[9] and gravure printing[10] have revealed a critical bottleneck presented by the upscaling of the charge transport layers, which are key for achieving high efficiency and long-term operational stability.[11] Inorganic charge transport layers such as $SnO_2$ and $NiO_x$ deliver long-term photostability[12] and enhanced thermomechanical reliability,[13] but two significant challenges remain for processing these materials. The first challenge is a lack of low capital expenditure (CapEx), scalable methods for high-speed deposition (10s - 100s m/min) of ultrathin (10 – 30 nm) inorganic CTLs in uniform, pinhole-free coatings over large areas (10– 1000's cm$^2$). Previous work on scalable HTL processing focused on slot-die coating at low speeds (< 1 m/min) using conductive polymers such as PEDOT:PSS[14] and spiro-OMeTAD[15] that have inferior thermomechanical stability.[16] Although inorganic HTLs have been demonstrated in mesoscopic cells by screen printing NiO pastes[17,18], these thick films are unsuitable for planar perovskite architectures capable of higher efficiencies. Finally, another major barrier to scaling is that these inorganic HTLs require slow, high temperature anneals to achieve high PCE[19]—temperatures above the thermal limits of polymer substrates and ITO electrodes. These challenges demand faster, more uniform, and more reliable scalable manufacturing of ultrathin inorganic CTLs to accelerate the commercial viability of planar cell architectures optimized for long-term stability.

This work develops a new scalable strategy based on high-speed flexography to meet the technological demand for fabricating ultrathin HTLs in planar inverted perovskite solar cells.

Flexography is advantageous for rapid coating of ultrathin wet films of electronic inks without encountering the *low-flow limit* in slot die coating[20] or the limits of droplet-wise deposition in methods such as spray pyrolysis.[13] This is essential for inorganic hole transport layers (HTLs) that must be characteristically thin (5 – 50 nm) to achieve high PCE,[21] otherwise requiring expensive vacuum methods such as atomic layer deposition (ALD).[22–24] As a ubiquitous graphic arts method, flexography carries lower capital expenditures (CapEx) than vacuum methods (ALD, evaporation, sputtering) because of the exceptionally high throughput (as high as 200 – 600 m/min).[25] Flexography also allows 2D patterning, an important feature for monolithic solar module integration and for enabling new high-efficiency back-contact architectures.[26] **Figure 1a** illustrates a comparison across charge transport layer coating methods, highlighting three features relevant to enhancing performance and scalability of perovskite technology: 1) High-speed deposition for reducing capital expenditures and enhancing throughput, 2) Ability to deposit uniform, ultrathin films on the 10 nm scale, and 3) 2D patterning for cell isolation and module integration. Roll-based printing methods such as flexography, offset printing, and gravure satisfy each of these requirements, although only flexography offers a combination of low-cost pattern carrying elements and the ability to print on both rigid and flexible substrates.

Here we present flexographic printed HTLs for enhancing the scalability of inverted perovskite solar cells utilizing a p-i-n architecture based on $NiO_x$ sol-gel inks. We demonstrate how these high-speed (60 m/min) methods can yield unmatched uniformity (8 Å variability) for ultrathin films over large areas (140 $cm^2$) with decreased pinhole density and increased transparency compared with spun or sprayed $NiO_x$. Engineering the precursor design of printed NiO inks allows the optimization of both the optoelectronic characteristics and surface chemistry to achieve high photovoltaic efficiency (> 15% PCE) perovskite solar cells while utilizing the highest reported speeds of any printing method yet applied to perovskite CTLs (**Table S1**).

## 2. Results & Discussion

In this work, we use flexographic printing to fabricate inverted perovskite solar cells with ultrathin $NiO_x$ HTLs, an optimal p-i-n architecture known for providing high efficiency and long-term stability.[11,27] Sol-gel based $NiO_x$ HTLs have significant advantages compared with the low cohesive fracture energy of nanoparticles and organic films,[13] which is important for long term thermomechanical stability. **Figure 1b** shows a schematic of the flexographic printing process, which can perform rapid large area printing and coating of various electronic inks. Flexography utilizes a blanket anilox roller, which is doctored by a steel blade to meter the ink volume before transferring ink to raised features on a photopolymer stamp, shown in **Figure 1c**. Careful design of the stamp and the inks themselves can allow for printing fine features at a resolution of less than 10 µm[28] while maintaining high uniformity over large areas. These printed films exhibit low areal pinhole densities of less than 0.45 pinholes/cm$^2$ as measured by large area scanning microscopy (**Figure S1**), which is significantly lower than the pinhole densities of spin coated (1-4 pinholes/cm$^2$) NiO$_x$ films.[13] **Figure 1d** shows NiO$_x$ films printed on rigid ITO coated glass substrates with 350 µm spacing between features and low, micron-scale line edge roughness (LER). The low LER of these printed isolated features could facilitate scaling of these gaps to well below 100 µm for the purpose of module integration. This patterning capability of flexography offers a direct benefit for monolithic perovskite solar module fabrication, eliminating complex scribing steps that add considerable CapEx and offering the ability for integration of both active layers, transport layers, and metal bus bars for module application.[29]

Our flexographic printing process patterns ultrathin NiO$_x$ films (5 – 20 nm) with high uniformity, showing less than 8 Å variation in a 10 nm thick film over an area of 140 cm$^2$ (**Figure 1e**). This is advantageous for PSC device performance because PSCs are sensitive to NiO$_x$ film thickness, with thicker films exhibiting higher optical absorption and increased series resistance, resulting in decreased efficiency.[30] Fabrication of highly uniform, ultrathin films is necessary to evade these limitations present in thicker NiO$_x$ films and is achievable through flexographic printing. Our high-speed (60 m/min) process can deposit highly uniform, ultrathin films on both flexible and rigid substrates, like the SiO$_2$ coated Si wafer shown in Figure 1e, allowing for integration of printed NiO$_x$ into fabrication of both flexible and tandem perovskite-silicon solar cell architectures. Other scalable printing methods, like gravure printing, risk damaging mechanically fragile Si substrates, limiting the applications to plastic substrates. Flexographic printing, in contrast, provides an ideal method for scalable manufacturing in roll-to-roll or sheet-fed processing, eliminating vacuum deposition steps which generally require longer fabrication times and methods such as spin coating, which are only viable at the lab scale.[31–34]

**2.1. NiO$_x$ Sol-Gel Ink Design and Printed Film Morphology**

Printing NiO$_x$ sol-gel inks requires precise precursor ink design to ensure that the resultant films are highly uniform and are of the intended thickness. In this work, we explore the coupling between the ink design and the flexographic printing process to understand how changes to the precursor impact the resulting film morphology. **Figure 1f** shows that the viscosity of the NiO$_x$ inks varies from 2.5 to 13.6 mPa·s over a range of 0.2 M to 2.0 M Ni(NO$_3$)$_2$ concentration for an applied shear rate in the range of 1100 – 8500 s$^{-1}$ consistent with shear rates during ink transfer.[35] Viscosity is increased with concentration due to the complexes formed between the metal nitrate salt and the organic solvent, 2-methoxyethanol (2-ME).[36] The viscosity of a flexographic ink influences the amount of ink transferred during printing as well as the uniformity of the film. Inks with higher viscosities have been shown to have increased transfer, but they also experience a stronger viscous fingering effect, or 'ribbing', which typically induces longitudinally oriented nonuniformities in the printed direction.[37,38] During flexographic printing, the largest effects from ribbing occur during ink transfer from the stamp to the substrate as the printing plate and substrate surfaces diverge beyond the nip, splitting the film of ink.[39]

Large area printing studies of the NiO$_x$ films reveal that high printing speed mitigates viscous fingering, leading to the most uniform HTL layers. **Figure S2** illustrates a comparison of contrast-enhanced large area scanning microscopy images obtained from prints at low and high print speeds as well as increasing viscosity. Lower speed (30 m/min) prints exhibit an isotropic, dapple-like, non-uniformity with a characteristic wavelength ($\lambda$) of approximately 300 μm, whereas high speed (60 m/min) prints level to a featureless film. These results match recent flexography studies with polymers[38], emphasizing a key benefit of flexography that transfer and uniformity improve as throughput is scaled up.

By varying the concentration of the ink from 0.5 M to 2.0 M, we are able to control the final dry film thickness from 6 – 20 nm (**Figure 1g**), a range that can be optimal for planar perovskite architectures. The thinnest films printed from the lowest viscosity NiO$_x$ inks produce the highest uniformity printed films at both low and high speeds (**Figure S2**). This can be understood by considering the leveling time ($\tau$) for surface tension ($\gamma$) to drive a film with thickness modulations over a wavelength ($\lambda$) to a uniform thickness ($h$), as expressed by Equation 1:

$$\tau = \frac{3\eta\lambda^4}{16\pi^4\gamma h^3} \quad (1)$$

This leveling time derived from the Navier-Stokes equation scales directly with ink viscosity ($\eta$) and modulation wavelength, but inversely with surface tension and thickness.[35,40] The inverse scaling of $\tau$ with thickness means that ultrathin films, like HTLs, require careful process design to achieve high uniformity. In this work, high density anilox rollers (25 μm cell size) are utilized to achieve ultrathin films while minimizing the wavelength of modulations, allowing for faster leveling.[41] Based on measured film thicknesses and modulation wavelengths, the inks used here are estimated to level in approximately 0.1 - 0.4s, significantly shorter than the observed drying times of approximately 2s. We note that it is possible to further decrease the ink viscosity to enhance leveling by changing the solvent ratio. For $NiO_x$ solutions mixed with ethanol (EtOH) and 2-ME, increasing the amount of EtOH used decreases the viscosity of the solution (**Figure S3**), which is to be expected since the viscosity of pure EtOH (1.1 mPa·s) is smaller than the viscosity of pure 2-ME (1.7 mPa·s).

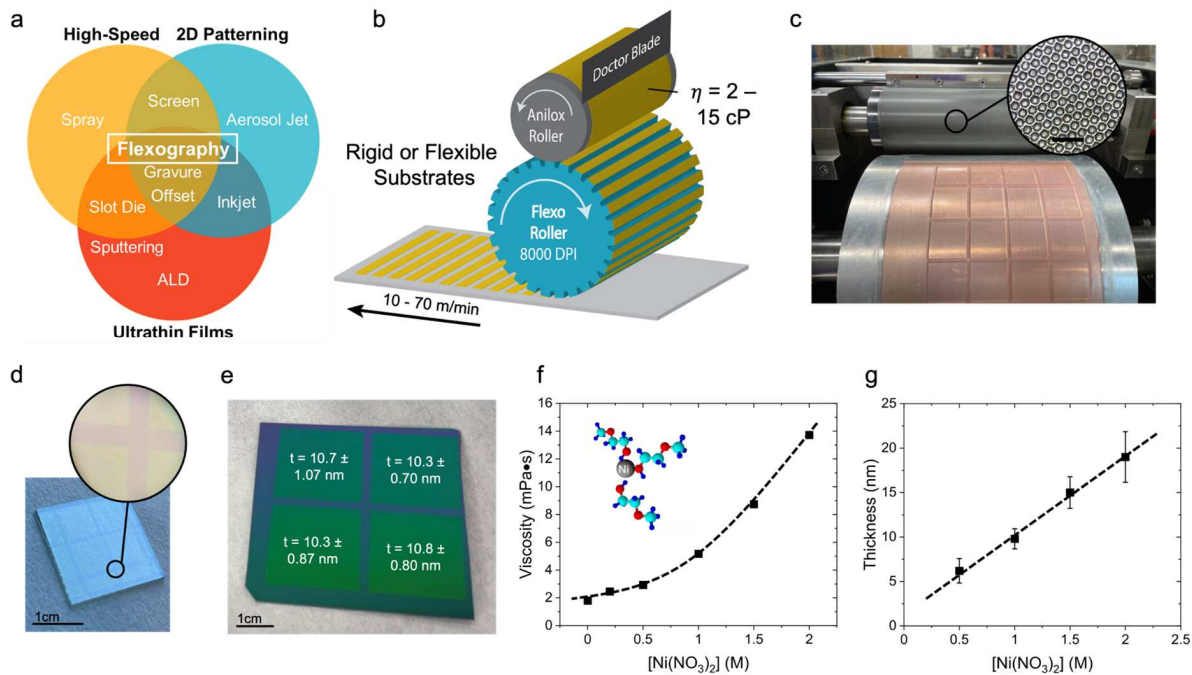

**Figure 1:** a) Comparison of thin film fabrication methods noting capabilities for high-speed, 2D patterning, and ultrathin film deposition. b) Scheme for flexographic printing indicating speeds and viscosities amenable to the process. c) Image of flexographic printer showing anilox roller and photopolymer stamp with inset displaying microscope image of engraved cells on the anilox roller (scale bar 250 µm). d) Flexographic printed $NiO_x$ pattern on ITO glass with inset showing low line edge roughness of printed $NiO_x$ (green) features. e) $NiO_x$ thin film flexographic printed on an $SiO_2$ coated Si wafer indicating thickness uniformity over a large area. f) Ink viscosity as a function of the $Ni(NO_3)_2$ concentration with inset showing complexes formed between the $Ni(NO_3)_2$ and 2-ME. g) Flexographic printed $NiO_x$ film thickness as a function of $Ni(NO_3)_2$ concentration.

### 2.2. Material Characterization of Printed $NiO_x$ Thin Films

Understanding the material properties of the printed films is vital to ensuring that the $NiO_x$ has the grain structure, chemical composition, and optical transmittance characteristic of a high performing HTL. Ultrathin printed $NiO_x$ films are observed to exhibit x-ray diffraction (XRD) spectra consistent with nanocrystalline cubic $NiO_x$, showing the characteristic primary (200) peak (**Figure 2a**). XRD spectra of printed films closely resemble those of spin-coated $NiO_x$, illustrating that the method for $NiO_x$ wet film deposition does not significantly impact the phase or crystallinity of the film. Although thicker spin coated films exhibit (200) peaks of higher intensity, detailed scans show a similar peak width (FWHM) for the (200) peak of both spin and printed films (**Figure S4**), suggesting comparable crystallite size (5.7 nm vs. 5.3 nm for printed vs. spin). Additionally, XPS of $NiO_x$ films indicates a similar composition to sputtered $NiO_x$.[42] High-resolution Ni 2p peaks are fit in **Figure 2b** to illustrate the chemical composition of the films, showing contributions from NiO (red), $Ni(OH)_2$ (blue), NiOOH (green), and $Ni^{3+}$ (purple) as well as satellite peaks having energies above 860 eV. The amount of NiO and $Ni^{3+}$ present in the printed films matches that of reported sputtered $NiO_x$ films, while the amount of $Ni(OH)_2$ in the printed films is slightly higher than in sputtered films due to increased surface hydroxide content resulting from the ultrathin nature of the printed $NiO_x$.[42,43] The same comparison can be made between printed $NiO_x$ films and thermally annealed or nanoparticle $NiO_x$ films, indicating that $NiO_x$ formed through flexographic printing provides very similar chemical composition to $NiO_x$ formed through other fabrication methods.[42] This evidence from XRD and XPS studies indicates that these ultrathin, flexographic printed $NiO_x$ HTLs can exhibit similar optoelectronic properties to high performing HTLs produced by various slower methods, ensuring th e potential for achieving high efficiency in inverted perovskite solar cells.

The flexographic printed $NiO_x$ films also have higher transmittance than typical spin coated $NiO_x$ since they can be made much thinner while maintaining conformal coverage. **Figure 2c** shows the transmittance spectra of $NiO_x$ with varying thickness for wavelengths in the range of 400 – 750 nm indicating a decrease in average transmittance from 94.9% to 93.3% as the thickness increases from 6 nm to 19 nm. Similarly, the absorbance spectra of the $NiO_x$ films measured with UV-VIS (**Figure S5**) show an increase in absorbance as the $NiO_x$ film thickness increases (**Figure 2d**). Integrated absorption (300 – 750 nm) shows a 36% increase for 19 nm films relative to 6 nm films. Additionally, compared to spin coated $NiO_x$, ultrathin printed $NiO_x$ films are less absorbing, showing improved light transmission for boosting the short circuit current density ($J_{sc}$) of inverted single junction PSCs as well as tandem architectures.

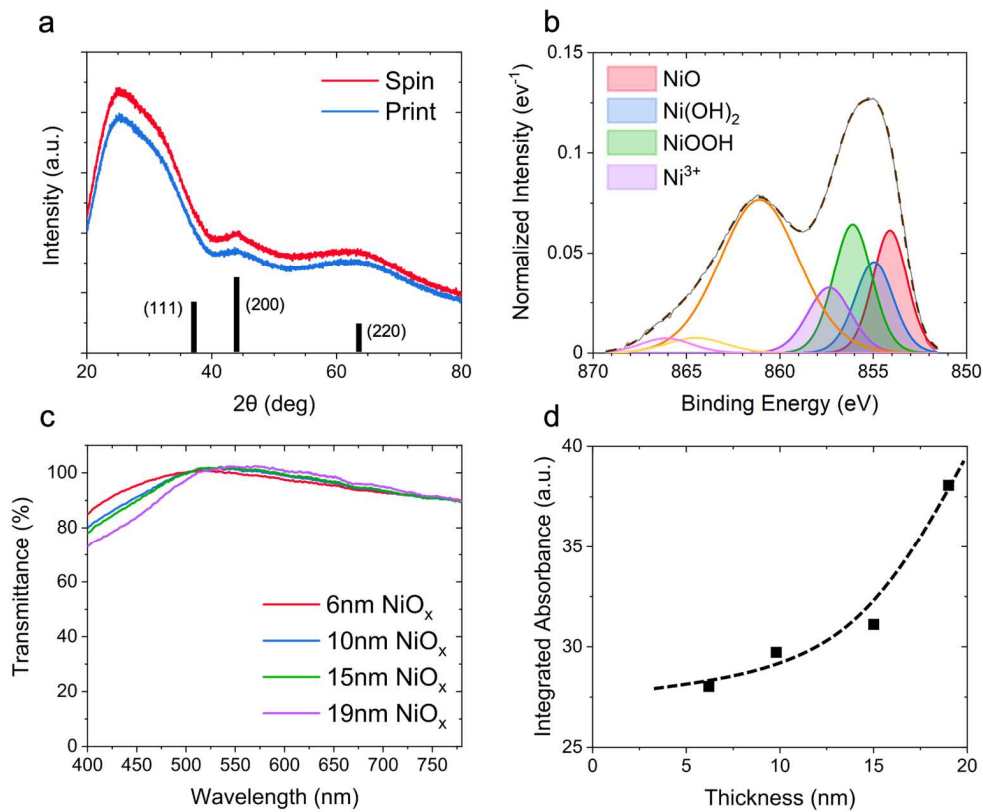

**Figure 2:** a) XRD spectra for spin coated and flexographic printed NiO$_x$ with reference peaks for FCC cubic NiO$_x$ (PDF Ref. No. 00-047-1049). b) XPS Ni 2p peak of flexographic printed NiO$_x$ with data shown in grey and fit shown in brown. c) Transmittance of flexographic printed NiO$_x$ films with varied thickness. d) Integrated absorbance (300 – 750 nm) of printed NiO$_x$ films.

## 2.3. Flexographic Printed NiO$_x$ Integration in Perovskite Solar Cells

These flexographic printed NiO$_x$ thin films were integrated as the HTLs in inverted, double cation perovskite solar cells (PSCs) to understand how the thickness and optoelectronic properties of printed NiO$_x$ films influence device performance. We utilize use a high-performance double cation (Cs$_{0.15}$FA$_{0.85}$PbI$_3$) perovskite, shown to improve device voltages and PCE,[42] with a C$_{60}$/BCP electron transport layer (ETL) architecture shown in **Figure 3a and 3b**. Implementing these printed HTLs in perovskite solar cells results in efficiencies that meet or exceed those of devices fabricated with similar precursors via other deposition methods, indicating that printed NiO$_x$ can achieve the continuous morphology and reliable optoelectronic properties necessary for implementation as a charge transport layer (CTL).[44] The smooth, wetting NiO$_x$ provides a uniform surface for perovskite film deposition with a very low contact angle compared to other common hole transport materials such as PTAA.[45] This allows for good crystallization of the perovskite on the NiO$_x$ surface (**Figure S6**), which can lead to homogenous films with larger grain sizes and fewer grain boundaries resulting in better device performance.[46] J-V curves for devices with spun and printed NiO$_x$ HTLs, as shown in **Figure 3c**, demonstrate this similar behavior for both PSCs. Printed devices achieve relatively high open circuit voltages ($V_{oc}$) of 1.04 V compared to other cells with a NiO$_x$ HTL. The printed devices achieve an average PCE of 13.1 ± 0.9 % with the champion device measuring 15.3% (**Figure S7**), whereas average spin coated NiO$_x$ yields devices with a PCE of 12.2 ± 2.0 %. The high short circuit current density (22.4 mA/cm$^2$) measured for these printed devices by J-V curves was confirmed by external quantum efficiency measurements (EQE) of integrated photocurrent (**Figure S8**). These enhancements could stem from higher visible range transmittance of the printed NiO$_x$, as discussed previously, since it is deposited in uniform films that are notably thinner than the 30 nm spin coated NiO$_x$.

The precursor ink design of the flexographic NiO$_x$ was varied to optimize the HTL film thickness specifically for integration into inverted planar double cation perovskite solar cells. **Figure 3d** shows how the NiO$_x$ film thickness impacts device performance. J-V curves for different thickness NiO$_x$ films indicate that thinner NiO$_x$ produces a higher $J_{sc}$ and exhibits higher efficiency. Thinner NiO$_x$ has higher transmittance, as discussed previously, resulting in increased light transmission through the material to the perovskite absorber. Perovskite films deposited on both printed and spin coated NiO$_x$ were measured by stylus profilometry, (**Figure S9**) indicating a similar average step height (480 nm), which suggests the change in $J_{sc}$ is not due to variations in the perovskite thickness. NiO$_x$ HTLs with a 10 nm thickness exhibited the highest printed device PCE (**Figure 3e**), exhibiting the lowest series resistance, resulting from the ultrathin feature of the printed layers. **Table S2** summarizes a comparison between the printed devices of varying NiOx thickness as well as the spin coated control devices. Devices with 10 nm NiO$_x$ outperform devices with thinner NiO$_x$ because they are thick enough to prevent shunting through the HTL while still conducting current efficiently (**Figure S10**). Cells with higher concentration NiO$_x$ prints (~ 20 nm) show an increase in series resistance of the devices (**Figure 3f**), further emphasizing the need for a process that can reliably deposit ultrathin $NiO_x$.

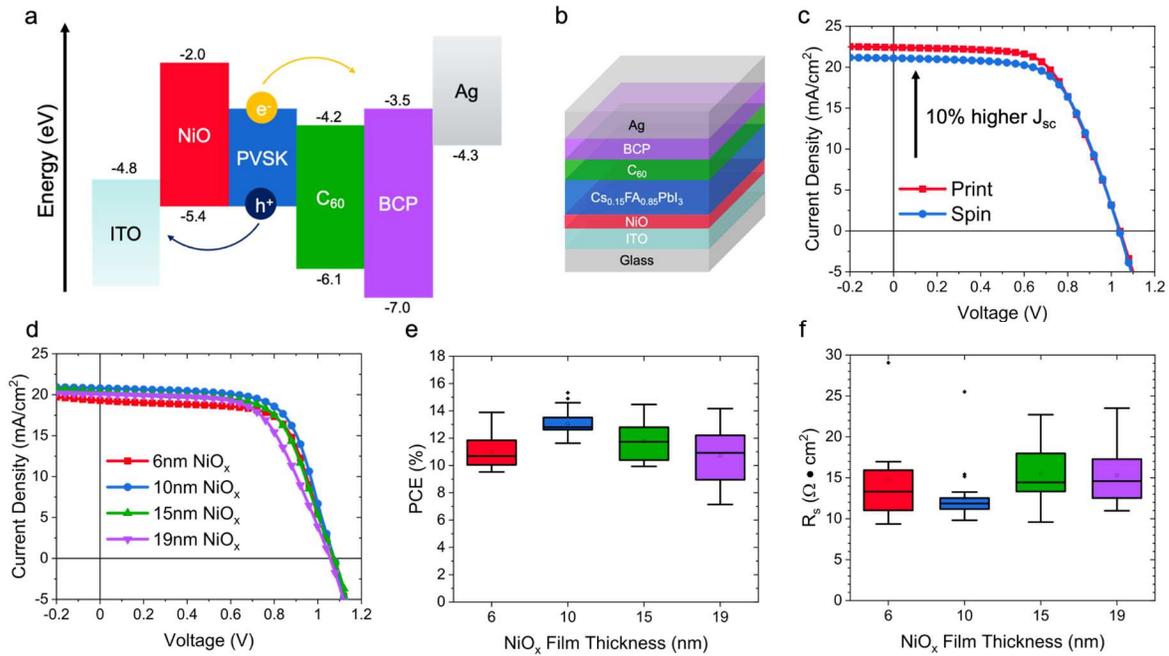

**Figure 3:** a) Energy diagram for perovskite solar cell (PSC) device architecture. b) Device architecture used for PSC fabrication. c) J-V characteristics for PSCs incorporating flexographic printed and spin coated $NiO_x$ films as an HTL. d) J-V characteristics for PSCs with printed $NiO_x$ HTLs of varied film thickness. e) PCE for PSCs with printed $NiO_x$ HTLs of varied film thickness. f) Series resistance for PSCs with printed $NiO_x$ HTLs of varied film thickness.

To further decrease the fabrication time for $NiO_x$ thin films and move towards a more scalable process, we explore rapid annealing of the $NiO_x$ films to understand how a faster post-anneal impacts the chemical composition. High resolution XPS O 1s scans of $NiO_x$ films annealed for 1 min and 60 min are shown in **Figure 4a** with a three-component fit of the spectra indicating contributions from NiO (529.5 eV), NiOOH (530.5 eV), and $Ni(OH)_2$ (531.3 eV) bonding states present in the films. Tracking the composition of the $NiO_x$ throughout the anneal is key to understanding the evolution of the optoelectronic properties of the material. During annealing, Ni-OH groups are converted to a Ni-O-Ni structure through condensation reactions that densify the film, allowing crystallization and the onset of electrical conductivity.[47] By monitoring the stoichiometry of the $NiO_x$, we can understand the progression of annealing towards the final state of the film. O 1s scans reveal that $Ni(OH)_2$ as well as NiOOH peaks exhibit slightly greater intensity in the rapid annealed films (1 min) as compared to the fully annealed films (60 min), both of which are dominated by surface hydroxide states rather than bulk hydroxide, a likely result of the ultrathin nature of these printed films.[43] The slight increase (~ 9%) in hydroxide for rapidly processed samples is to be expected since a shorter anneal allows less time for densification and conversion to an Ni-O-Ni structure from an Ni-OH dominated structure. Increased hydroxide content has been shown to significantly impact the electrical properties of NiO, notably reducing the resistivity and decreasing transmittance.[48] However, the surface hydroxide content of the printed films did not increase significantly as the anneal time increased, which is consistent with the small (2X) change in resistivity between the rapidly and fully annealed films shown in **Figure 4b**. Additionally, the resistivity of the printed $NiO_x$ films in this study matches that of $NiO_x$ fabricated through other scalable methods, like spray coating,[13,49] further indicating that the slightly elevated hydroxide content in the printed films does not limit the overall performance of the film as an HTL.

Perovskite cells incorporating flexographic printed ultrathin $NiO_x$ reached high performance even with rapid annealing (1 min) when compared with devices utilizing longer annealing steps (60 min). All printed $NiO_x$ films exhibit high visible range transmittance, averaging 95 – 97% transmittance over a range of 400 – 780 nm, as shown in Figure 4b, with rapid annealed films exhibiting high transmittance (> 97%) over a larger range of wavelengths than fully annealed films (**Figure S11**). This leads to high $J_{sc}$ and improved overall performance of double cation perovskite cells, as shown in **Figure 4c**. J-V characteristics for cells with these $NiO_x$ films indicate that rapid annealing can achieve over 90% of fully annealed PCE with 60X shorter total fabrication time. Cells incorporating printed $NiO_x$ can outperform devices with spin coated $NiO_x$ because they are incredibly thin (5 – 8 nm). Our method allows for uniform film deposition at thicknesses previously only achievable through ALD (2.5 – 7.5 nm),[30] however, ALD requires high vacuum and long deposition times, making it incompatible with module scaling. Flexographic printing provides a high-speed alternative for ultrathin $NiO_x$ deposition over large areas, resulting in films with tunable thickness and high transmittance and allowing for fabrication of perovskite cells with higher PCE and $J_{sc}$ without the CapEx required for ALD.

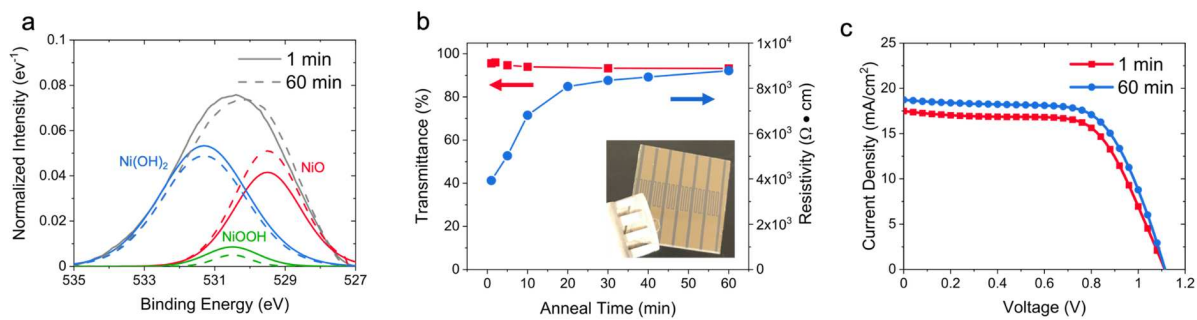

**Figure 4:** a) XPS O 1s spectrum of flexographic printed $NiO_x$ films annealed for 1 min (solid) and 60 min (dashed). b) Average visible range transmittance and measured resistivity of printed $NiO_x$ films annealed for 1 – 60 min with inset showing conductivity structure used for resistivity measurement. c) J-V characteristics of PSCs with 1 min and 60 min annealed printed $NiO_x$ HTLs.

By combining flexography with a rapid thermal anneal, we have shown that it is possible to reduce the total processing time for $NiO_x$ thin films by over 60X compared with reports of thicker spin coated $NiO_x$ films.[27] Flexographic printed $NiO_x$ films also have the potential to be integrated with photonic annealing, a fast, reliable method for rapid thin film annealing.[50] However, photonic annealing requires drying films prior to annealing, elongating the total fabrication time despite its quick nature.[50] With wet films approximately 1 µm in thickness and drying times of only 1 – 2 s at room temperature, flexographic printed $NiO_x$ is a good candidate for photonic annealing, a process that could alleviate the dependence on high-temperature thermal annealing that is currently used for $NiO_x$ thin film fabrication, and allow fabrication of flexible inorganic device architectures on lower cost polymer substrates. In combination with methods recently developed to print PCBM / BCP ETL layers,[51] our flexographic HTLs can provide a platform for fully printed perovskite cell architectures.

## 3. Conclusion

In summary, we present a method for rapid fabrication of ultrathin $NiO_x$ to enhance scalability and boost the performance of planar perovskite solar cells. This process employs low CapEx flexographic printing to accelerate $NiO_x$ deposition to 60 m/min, significantly beyond the state of the art of printed HTL layers, while patterning pinhole-free films with sub-nm thickness control over areas larger than 140 $cm^2$. Engineering the $NiO_x$ inks allows the optimization of the optoelectronic characteristics and surface chemistry to achieve high efficiency (> 15% PCE) PSCs while utilizing the fastest reported speeds of any printing method applied to perovskite CTLs. The compatibility of this process with both flexible and rigid substrates promises to unlock the potential for high-speed fabrication of both single junction PSCs as well as silicon-perovskite tandems, eliminating the manufacturing bottleneck induced by slow deposition of charge transport layers.

## 4. Experimental Section

*Precursor Preparation*

Spin coated NiO$_x$ inks were formulated from a 1.0 M Ni(NO$_3$)$_2$·6(H$_2$O) (99.9985% Alfa Aesar) solution in ethylene glycol (EG) with 1.0 M ethylenediamine (EDA). Printed NiO$_x$ inks were formulated with multiple solvent ratios and molarities to optimize their performance in perovskite solar cells. Printed films used for the NiO$_x$ thickness study implemented a combustion solution consisting of a Ni(NO$_3$)$_2$·6(H$_2$O) solute dissolved in 2-ME with 1.0 M acetylacetone added as fuel. Printed films used for the timed anneal study consisted of a 0.2 M Ni(NO$_3$)$_2$·6(H$_2$O) combustion solution in 1 mL of a 1:1 ratio of 2-ME to ethanol with 0.02 g acetylacetone as fuel. The double cation perovskite solution consisted of 0.0546 g CsI (>99.0% TCI), 0.2108 g FAI (GreatCellSolar Materials), and 0.6454 g PbI$_2$ (99.9985% Alfa Aesar) dissolved in 2 mL of solvent with a 3:1 v/v ratio of DMF:DMSO. Anhydrous DMF and DMSO were both obtained from ACROS Organics and stored in an inert nitrogen glovebox where the perovskite solution was also prepared and stored. Once mixed, the perovskite solution was stirred overnight at 65°C to dissolve. All precursors were used within two days of mixing. 99.9% C$_{60}$ was obtained from Fisher Scientific and >99.0% bathocuproine (BCP) was obtained from TCI, both were used for thermal evaporation.

*Flexographic Printing*

Flexographic printing was done with a 1200 line per inch (LPI) ceramic coated anilox roller (~ 20 μm cell size) at a speed of 60 m/min utilizing a 200 mm x 250 mm flexographic printing plate. ITO coated glass substrates (2 cm x 2 cm, 15 Ω sqr$^{-1}$, Xin Yan Technology, Ltd.) were cleaned by sequential sonication in an alkaline detergent solution (Extran), water, acetone, and isopropyl alcohol. They were subsequently dried and treated with UV-ozone for 20 min at 25 mW cm$^{-2}$. SiO$_2$ substrates were cleaned with air plasma (*Plasmaetch*) for 1 min at 1000 W. 75 – 150 μL of solution was deposited on the anilox roller and allowed to spread out for 2 s before printing. Substrates were secured in place via Kapton tape, which was placed on the edge of the substrate, overlapping about 2 mm at the edge the surface. Flexography printing was conducted at 60 m/min with line contact pressures of 40 – 80 kPa between the flexographic printing plate and the substrate. The substrates were then immediately transferred to a hotplate where they were annealed in air for 60 min (unless otherwise stated) at 300°C. The anilox roller, stamp, and doctor blade were cleaned with IPA and allowed to dry completely between prints. Printed films' uniformity and continuity (pinhole densities) were characterized by using large area scanning microscopy (Keyence VHX-7000) to stitch high-resolution micrographs over ~ 10 cm$^2$.

*Solar Cell Fabrication and Characterization*

NiO$_x$ was deposited on cleaned ITO either by spin coating 20 µL of solution at 5000 rpm for 32 s or flexographic printing as described above. Following deposition, the NiO$_x$ films were annealed at 300°C for 60 min (unless otherwise stated) and then moved into an inert nitrogen glovebox for perovskite deposition. Spin coated NiO$_x$ films had a thickness of 30 nm while printed NiO$_x$ films had a thickness of 5 – 20 nm depending on the ink used for printing. 20 µL of the perovskite solution was spread onto the substrates and spin coated at 5000 rpm for 50 s. When there were 25 s remaining in the spin, 300 µL of chlorobenzene (CB) antisolvent was deposited. After spinning, the devices were transferred to a hotplate where they were annealed for 40 min at 100°C. Following the perovskite anneal, the samples were cooled to room temperature and completed with an electron transport layer and back electrode consisting of a 30 nm layer of either spin coated PCBM or evaporated C$_{60}$ followed by a 5 nm layer of BCP and a 100 nm thick Ag metal electrode.

The perovskite cells were measured in ambient conditions (≈ 45% RH, 25 °C) under 1 sun, AM 1.5G illumination (Oriel LSH-7320 Solar Simulator). The devices were fabricated with a pixel area of 0.134 cm$^2$. The lamp intensity was set based on an NREL calibrated Si reference cell. J-V curves were collected with a precision sourcemeter (B2902A) measured between –0.2 and 1.2 V with an increment of 0.01 V and a delay of 0.06 s between points. External Quantum Efficiency (EQE) measurements were obtained using a calibrated reference photodiode (Newport, Power Meter 843-R with detector model 818-UV) and a monochromator (Optometrics Model DMC1-03) set to a FWHM of 5 nm. EQE photocurrent measurements were collected using a broadband light source from 400 to 890 nm at wavelength increments of 10 nm from 300 nm to 890 nm. UV-A range EQE were obtained using multiple UV-LED sources (*UVTOP LEDs*) to cover wavelengths from 300 – 410 nm.

*Material Characterization*

X-ray diffraction (XRD) analysis was performed on NiO$_x$ films deposited on soda-lime glass using a Rigaku MiniFlex diffractometer with Cu Kα radiation at a voltage/current of 40 kV/15 mA with step size of 0.02° at scanning rates of 1° and 0.1° per min. SEM analysis was performed with a Thermo Fisher Scientific Helios 5 CX DualBeam SEM. X-ray photoelectron spectroscopy (XPS) analysis was performed with a Kratos Axis Supra. Viscosity was measured with a micoVISC-m at a shear rate between 1100 and 8500 $s^{-1}$. Transmittance was measured with a Vernier SpectroVIS Plus Direct Spectrometer in transmittance mode. UV-VIS absorption measurements were collected with a Denovix DS-11 FX Plus. $NiO_x$ film resistivity was measured using interdigitated test structures probed with a semiconductor parameter analyzer (Agilent E5260A). Film thicknesses were measured with an AlphaStep D-500 stylus profilometer with a probe weight of 1.0 mg and a scan speed of 15 mm/s.

**Supporting Information**

Supporting Information is available from the author.

**Acknowledgements**

The manuscript was written through contributions of all authors. All authors have given approval to the final version of the manuscript. This work was sponsored by a seed grant from the Arthur L. Irving Institute for Energy and Society at Dartmouth. We acknowledge John Wilderman at the University of New Hampshire for his help in completing XPS measurements.

Supporting Information

**Eliminating the Perovskite Solar Cell Manufacturing Bottleneck via High-Speed Flexography**

*Julia E. Huddy, Youxiong Ye, and William J. Scheideler\**

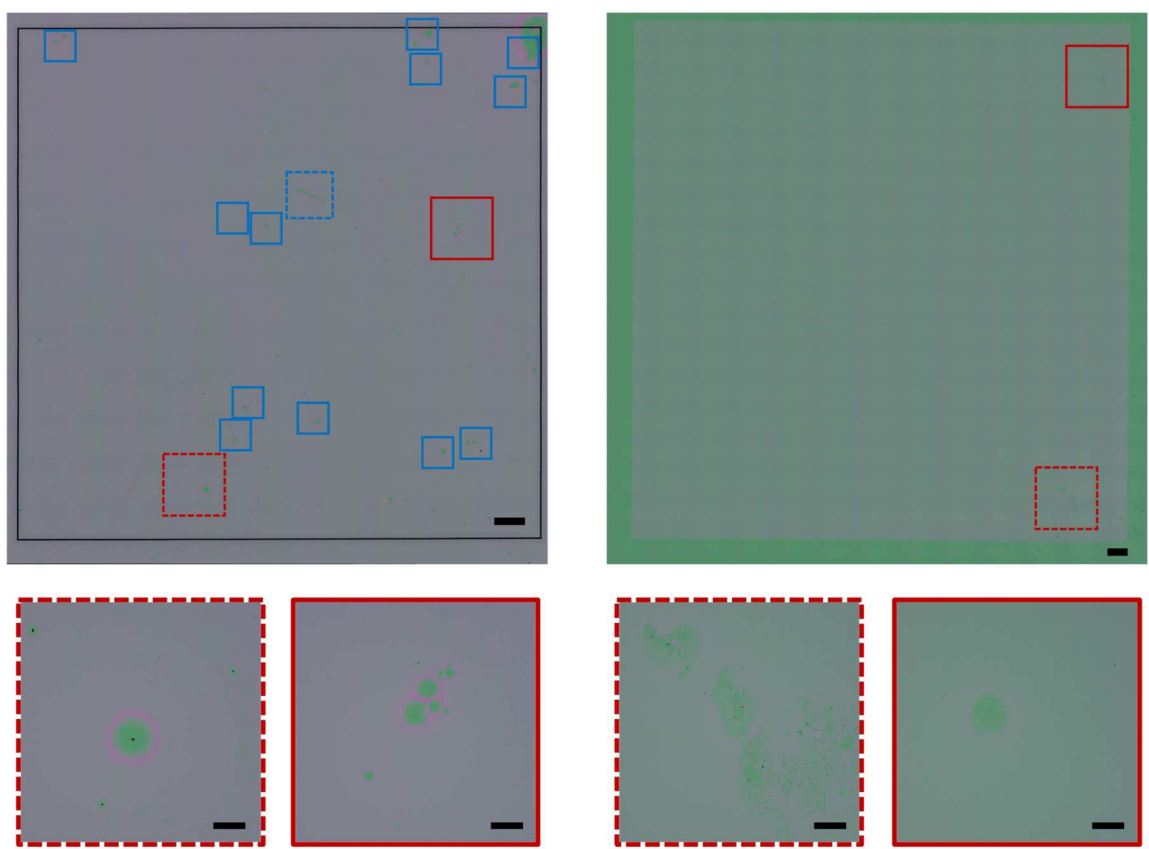

**SI Figure 1:** Microscope images of spin coated (left) and printed (right) NiO$_x$ on SiO$_2$ showing pinhole densities of 3.7 pinholes/cm$^2$ and 0.45 pinholes/cm$^2$ respectively. Blue boxes indicate areas with one or more pinholes. Red boxes indicate areas matching high resolution insets shown below. Scale bars are 1 mm for the top images and 250 µm for the bottom inset images.

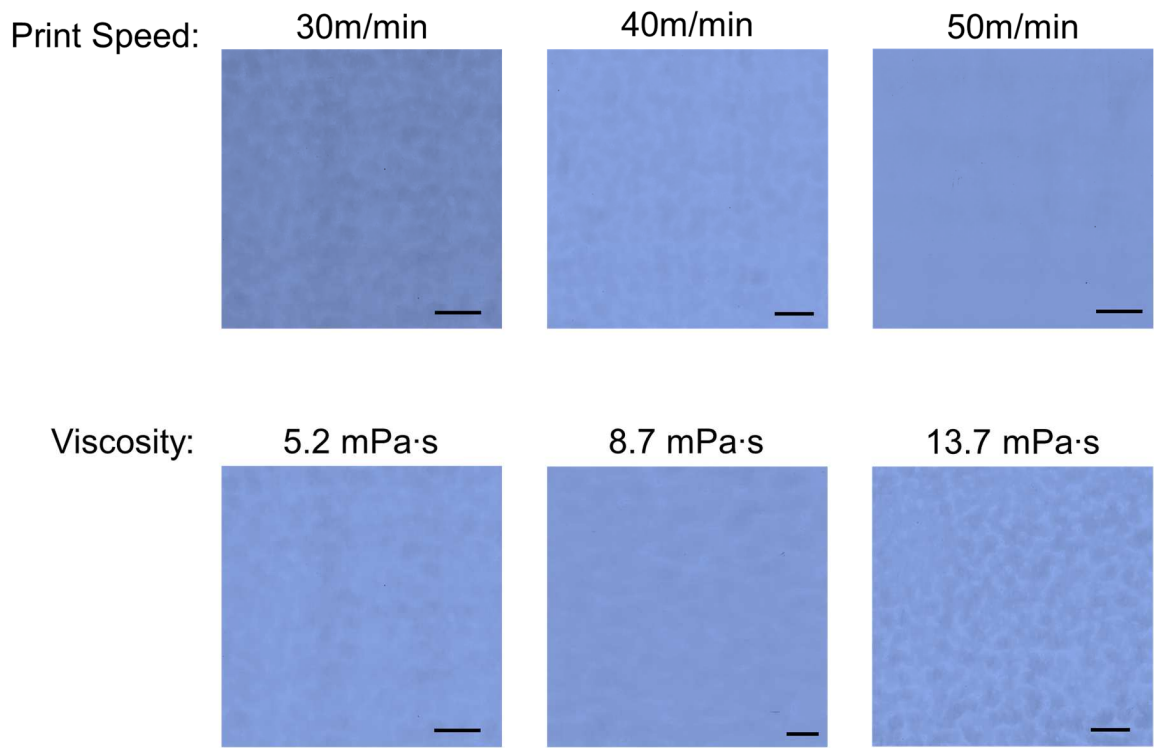

**SI Figure 2:** Top row shows large area, contrast-enhanced images of printed NiO$_x$ films at varying speeds with 1.0M NiO$_x$ ink. Bottom row shows low speed (30 m/min) printed films with varying viscosity (1.0M, 1.5M, 2M concentrations). Scale bar is 1mm in each image.

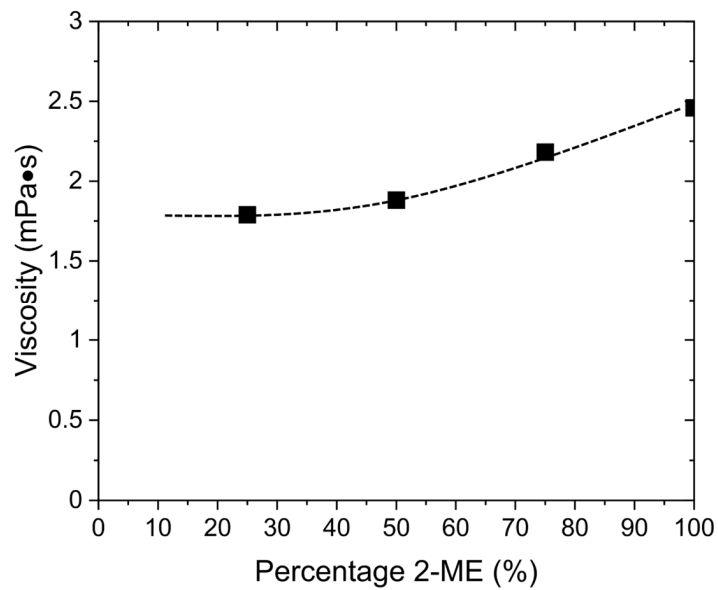

**SI Figure 3:** Viscosity of 0.2 M Ni(NO$_3$)$_2$ solutions in varied volumetric ratios of 2-ME to EtOH.

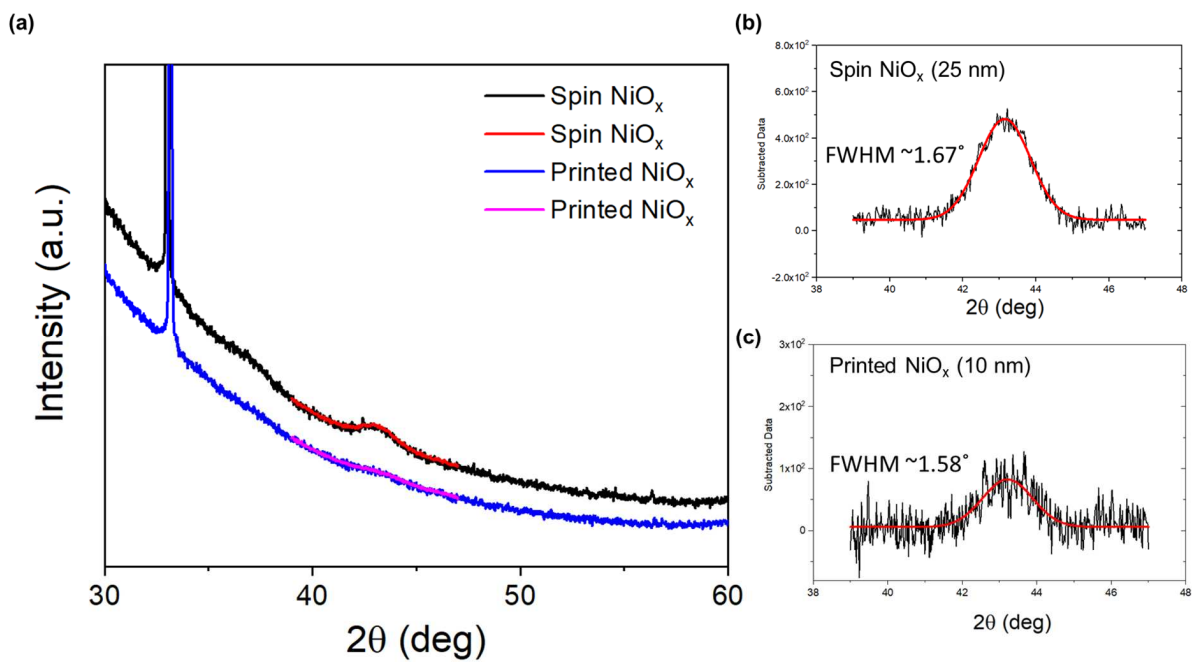

**SI Figure 4:** (a) Broad (black, blue) and detailed (red, pink) scans of NiO$_x$ films spin coated or printed on Si substrates. (b) Detailed scan with Gaussian peak fit for (200) peak for spin coated NiO$_x$ film. (c) Detailed scan with Gaussian peak fit for (200) peak for printed NiO$_x$ film.

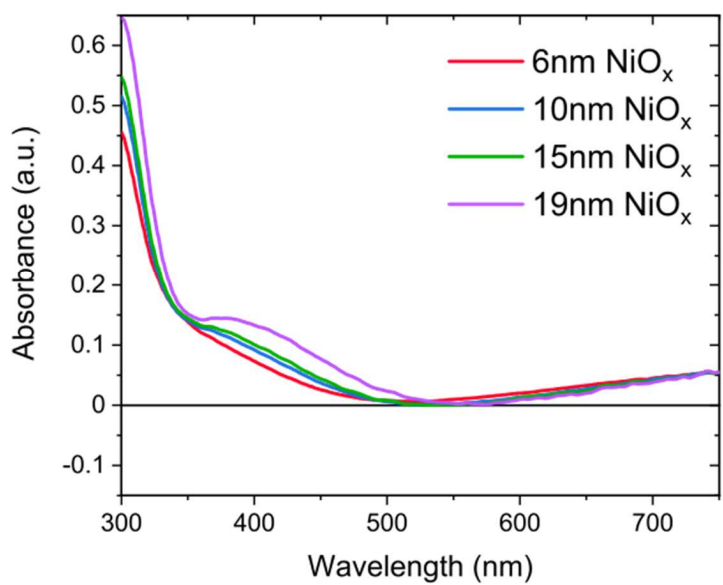

**SI Figure 5:** UV-VIS spectra used to determine absorbance constant.

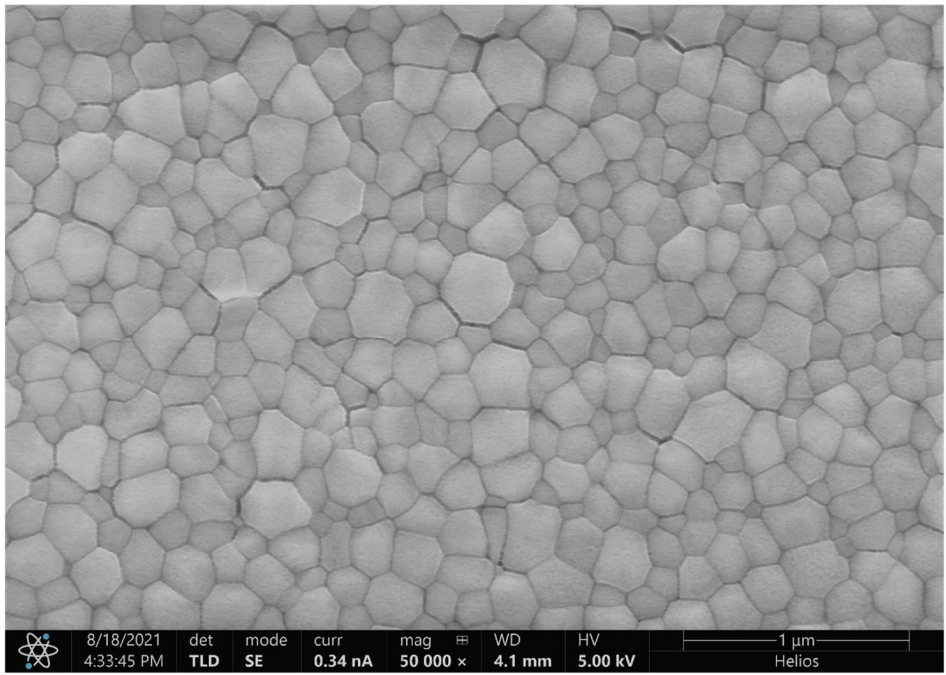

**SI Figure 6:** SEM image showing perovskite grain structure as deposited on flexographic printed NiO$_x$.

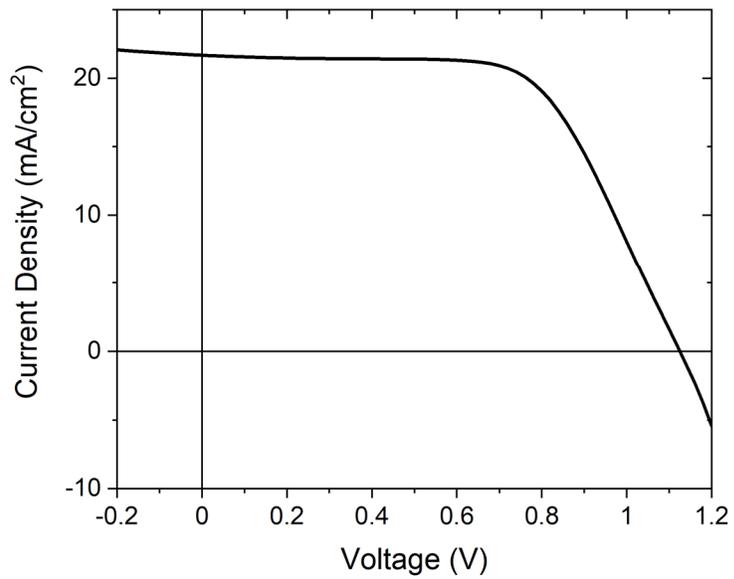

**SI Figure 7:** Champion perovskite solar cell with printed NiO$_x$ HTL (PCE of 15.3%, V$_{oc}$ of 1.12 V).

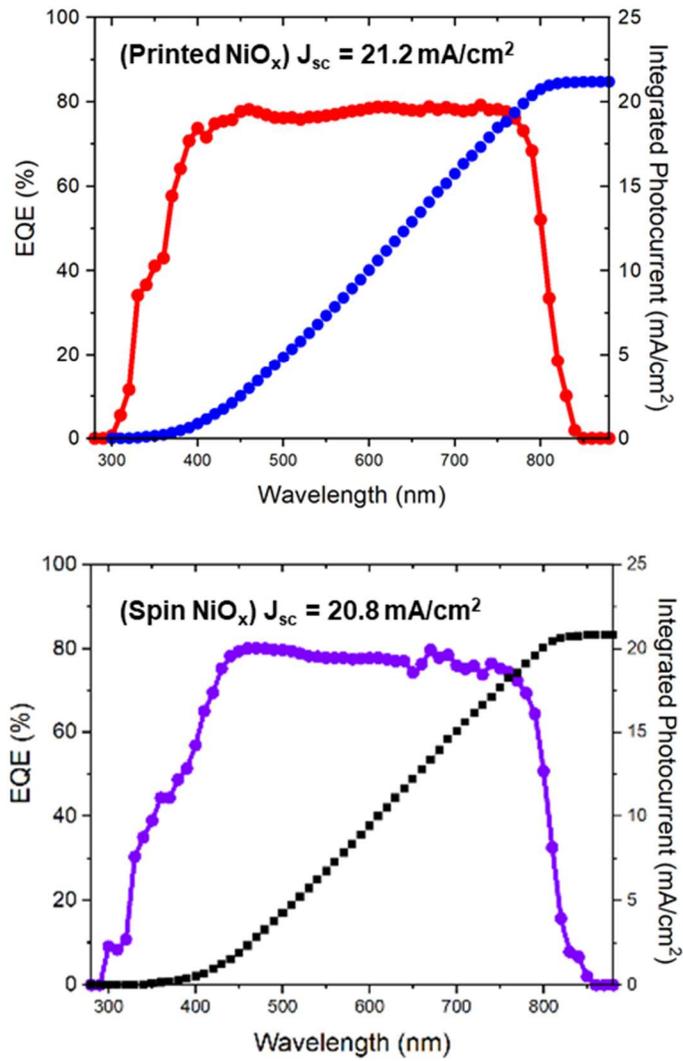

**SI Figure 8:** Measured external quantum efficiency (EQE) of perovskite solar cell with printed $NiO_x$ HTL (top) and spin coated $NiO_x$ (bottom) with integrated short circuit current shown on right axis.

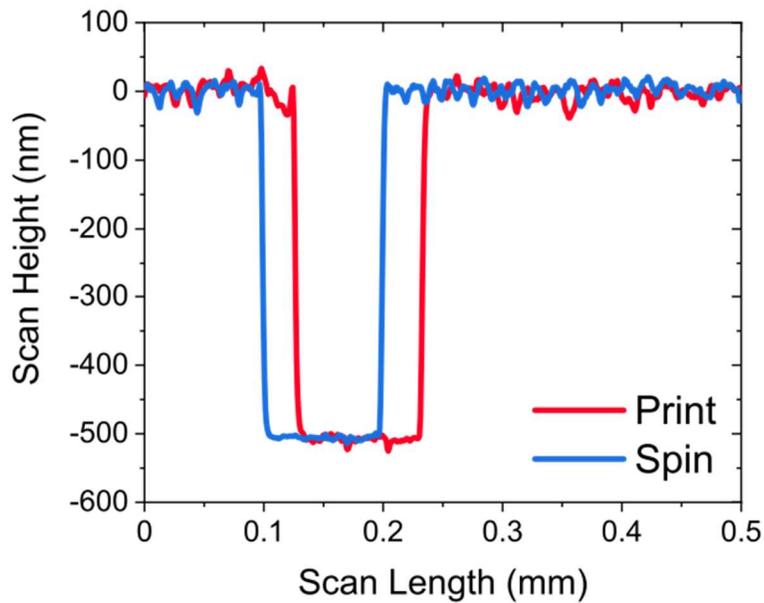

**SI Figure 9:** Stylus profilometry scans showing combined perovskite/ETL thickness as deposited on printed and spun $NiO_x$.

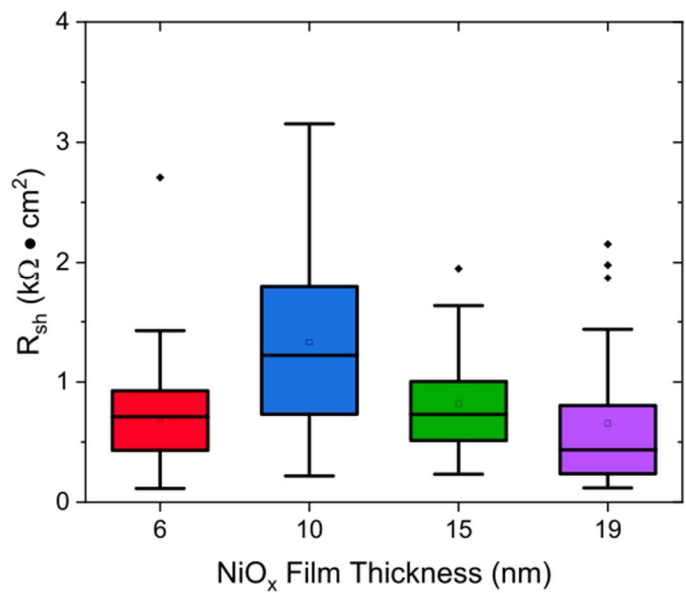

**SI Figure 10:** Shunt resistance for PSCs with flexographic printed $NiO_x$ HTLs of varied film thickness.

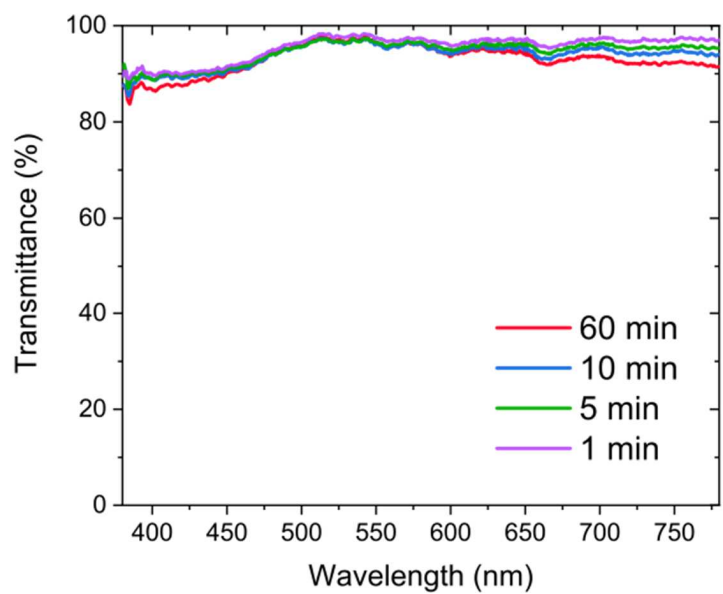

**SI Figure 11:** Transmittance spectra for flexographic printed NiO$_x$ with varied anneal time as measured on ITO coated glass.

**Table S1** – Comparison of reported scalable printed HTL materials for perovskite solar cells.

| Author | Year | HTL | Method | Web Speed | PCE % (Average, *Peak) | Ref |
|---|---|---|---|---|---|---|
| **Schackmar et al.** | 2021 | NiO | Inkjet | N/A | 12.6 (*17.2) | [51] |
| **Bashir, et al.** | 2019 | NiO | Screen | N/A | *13.7 | [18] |
| **Cao, et al.** | 2015 | NiO | Screen | N/A | 13.4 (*15.0) | [17] |
| **Scheideler, et al.** | 2019 | NiO | Spray | 1.5 m / min | 16.2 (*17.7) | [13] |
| **Y.Y. Kim, et al.** | 2019 | Spiro-OMeTAD | Gravure | 18 m /min | 16.0 (*17.2) | [10] |
| **B. Dou, et al.** | 2018 | Spiro-OMeTAD | Slot-Die | 0.5 m / min | *14.1 | [15] |
| **T.M. Schmidt, et al.** | 2015 | PEDOT | Slot-Die | 0.5 m / min | *9.4 | [14] |

**Table S2** – Comparison of device characteristics for solar cells with printed $NiO_x$ of varied thickness and spin coated $NiO_x$. Including an integrated $J_{sc}$ calculated from EQE measurements. *indicates J-V scan $J_{sc}$ measurement for devices measured by EQE.

| $NiO_x$ Condition | $V_{oc}$ (V) | $J_{sc}$ (mA/cm$^2$) | FF (%) | PCE (%) | Max PCE (%) | Integrated $J_{SC}$ (mA/cm$^2$) | N |
|---|---|---|---|---|---|---|---|
| Print (6 nm) | 1.00 ± 0.03 | 19.4 ± 1.4 | 56.5 ± 4.5 | 11.0 ± 1.1 | 13.9 | | 26 |
| Print (10 nm) | 1.02 ± 0.04 | 21.3 ± 1.4 | 60.1 ± 2.9 | 13.1 ± 0.9 | 15.3 | 21.2 (*21.6) | 26 |
| Print (15nm) | 1.03 ± 0.04 | 20.7 ± 1.3 | 55.4 ± 5.6 | 11.8 ± 1.4 | 14.5 | | 26 |
| Print (19 nm) | 1.02 ± 0.04 | 19.6 ± 1.6 | 53.6 ± 7.4 | 10.7 ± 1.9 | 14.2 | | 26 |
| Spin | 1.01 ± 0.04 | 21.3 ± 1.6 | 57.0 ± 8.1 | 12.2 ± 2.0 | 16.3 | 20.8 (*21.2) | 55 |